# Improved current saturation and shifted switching threshold voltage in In$_2$O$_3$ nanowire based, fully transparent NMOS inverters via femtosecond laser annealing

Chunghun Lee[1], Sangphill Park[1], Pornsak Srisungsitthisunti[1], Seongmin Kim[1], Chongwu Zhou[2], David B. Janes[1], Xianfan Xu[1], Kaushik Roy[1], Sanghyun Ju[3], and Minghao Qi[1]

[1]*School of Electrical and Computer Engineering, and Birck Nanotechnology Centre, Purdue University, 465 Northwestern Avenue, West Lafayette, IN 47907, USA*

[2]*Department of Electrical Engineering, University of Southern California, 3710 McClintock Avenue, Los Angeles, California 90089, USA*

[3]*Department of Physics, Kyonggi University, Suwon, Kyeonggi-Do 443-760, Republic of Korea*

**Transistors based on various types of non-silicon nanowires have shown great potential for a variety of applications, especially for those require transparency and low-temperature substrates. However, critical requirements for circuit functionality such as saturated source-drain current, and matched threshold voltages of individual nanowire transistors in a way that is compatible with low temperature substrates, have not been achieved. Here we show that femtosecond laser pulses can anneal individual transistors based on In$_2$O$_3$ nanowires, improve the saturation of the source-drain current, and permanently shift the threshold voltage to the positive direction. We applied this technique and successfully shifted the switching threshold voltages of NMOS based inverters and improved their noise margin, in both depletion and enhancement modes. Our demonstration provides a method to trim the parameters of individual nanowire transistors, and**

**suggests potential for large-scale integration of nanowire-based circuit blocks and systems.**

Flexible and/or transparent electronics have attracted significant interests due to their potential applications including see-through, light-weight and conformable product[1,2,3,4,5]. In particular, nanowire transistors (NWTs) may be better suited for future display products requiring transparent electronic switches because NWTs offer higher carrier mobility than those of thin-film transistors (TFTs), as well as the low-temperature processes that are compatible with optical transparency requirements[2,3,6]. High performance NWTs typically uses ZnO, $SnO_2$, and $In_2O_3$ semiconducting oxide nanowires, or aligned/random networked single-walled carbon nanotubes[1,2,4,6,7]. Many reports have suggested that NWTs have higher performance and more stable transistor characteristics compared with amorphous silicon and poly-silicon TFTs, especially for field effect mobility ($\mu_{eff}$) and subthreshold slope (SS)[8,9,10,11]. Despite these excellent properties (high performance, high sensitivity and high efficiency), however, there are still many issues to be resolved before NWTs can find practical applications. Some of the most urgent ones for digital and analog applications are highly saturated transistor current and robust semiconductor characteristics, such as uniform and controllable threshold voltages ($V_{th}$) and SS. Even though many un-passivated NWTs have been demonstrated, source-drain currents are not saturated but rather increase slightly linearly in most reports[2,4,7,8,9,10,11,12]. Little research, to our knowledge, has been conducted to reduce such linear increase even though it is perhaps the biggest obstacle for the incorporation of NWTs in such transparent circuitry on low temperature substrates, as current saturation is the key benefit of transistors. While high-temperature annealing or doping could be used to mitigate this problem in commercial thin-film transistors, elevated temperatures can change the properties of semiconducting nanowires, and there are difficulties in adjusting the doping level uniformly. Furthermore, these methods are in most cases incompatible with flexible device panels.

Here we report the effects of femtosecond laser annealing on fully transparent inverters consisting of two $In_2O_3$ NWTs, and show that their current saturation is improved (3-7 times increase in output resistance) and the inverting voltages can be permanently shifted. Focused laser annealing is useful in that it can be applied selectively to small areas that require high temperatures. As a result, component damages at the time of conventional thermal annealing of the entire panel can be avoided and unwanted effects in those areas could be excluded from the annealing process[13,14]. In our process, we focused the laser beam spot at the contact area rather than on the nanowires themselves to avoid damaging or sputtering away them (Fig. 1a). Furthermore, this annealing process could be possible even on plastic panels because instantaneous laser annealing, which is performed on a length scale of several micrometers, does not affect the temperature of the entire panel. Using this method, we demonstrated switching threshold voltage control in fully transparent NMOS inverters with the load being a diode connected n-type $In_2O_3$ NW transistor operated in both the enhanced mode and depletion mode.

**Femtosecond laser annealing and its effect on $In_2O_3$ nanowire transistors**

Figure 1a is a cross-sectional view of the fully transparent NWT with the bottom gate structure, consisting of transparent glass substrate (corning glass), a buffer layer of 100nm thick silicon dioxide, a gate electrode made from 110nm thick patterned indium-tin oxide (ITO), a 20nm thick $Al_2O_3$ gate insulator through atomic layer deposition (ALD), a single-crystal semiconducting $In_2O_3$ nanowire as the active channel, and 110nm thick ITO for source/drain (S/D) electrodes. $In_2O_3$ nanowires were synthesized through a laser ablation method (band gap $E_g \sim 3.6eV$, and diameter $D \sim 20$ nm)[15]. They are transparent to visible light, and are suitable for transparent and flexible TFTs. Meanwhile, ITO is a promising candidate as transparent conductors for gate, source and drain electrodes[16,17,18] in TFTs. Figure 1b shows the field emission scanning electron microscope (FE-SEM) image of several NWT devices including all transparent components. The lengths of single $In_2O_3$ nanowire (~20nm diameter) addressed between S/D electrodes were ~3 μm to avoid the complications of the short channel effects. The thickness of the ALD $Al_2O_3$ gate insulator ($t_{ox}$) was ~30 nm. High-*k* $Al_2O_3$ gate dielectric showed excellent insulating properties, with an electrical breakdown field of > 8 MV/cm and a dielectric constant of ~9[19].

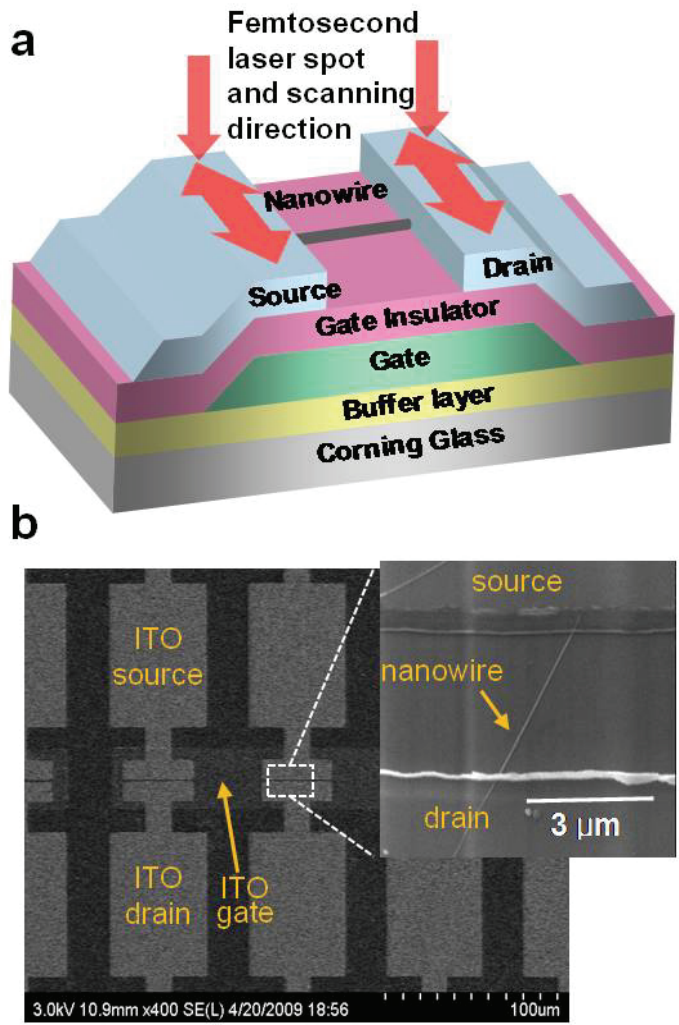

Figure 1 | **Schematic and scanning-electron micrograph of an $In_2O_3$ based NWT. a**, The cross-sectional schematic of a fully transparent, bottom gated nanowire transistor. The femtosecond laser pulses focus on the ITO source and drain area and scans along the edge of the source and drain pads. Laser pulses

5do not scan across the channel of the transistor, or the exposed portion of the nanowire. **b,** Top-view scanning-electron micrograph of a fully transparent NWT. ITO was used for gate, source, and drain. The inset shows a single $In_2O_3$ nanowire (D/L ~ 20 nm/3 μm) addressed between source and drain.

Figure 1a also illustrates the femtosecond laser annealing process. The unique aspect of our annealing process was that laser pulses were only focused on and scanned along the source/drain (S/D) contact regions using its particular property of localized energy input (beam spot diameter ~1.22 μm). If focused and scanned across directly on the nanowire region, however, the nanowire would have a high chance to be sputtered away and the transistor would be destroyed. The laser pulse duration was 50 femtosecond (fs) and the repetition rate was 1 kHz. Laser transmitted power varied from 1.67 μW (average energy fluence rate of 0.14 $J/cm^2$/pulse) to 5 μW (average energy fluence rate of 0.43 $J/cm^2$/pulse). The pulse wavelengths were centred at 800 nm, which has energy below the band gap of $In_2O_3$. Therefore we expected the effect to be likely different from the annealing using excimer lasers[13], which has a photon energy above the band gap of the nanowire.

The most prominent effects of laser annealing were the improvement of the current saturation and the positive shift of the threshold voltage $V_{th}$. Figure 2a shows the drain current versus drain-to-source voltage ($I_{ds}$–$V_{ds}$) characteristics for a representative NWT with $V_{gs}$ ranges from -1.5 V to 4 V in 0.5 V steps before (black open square) and after (red open circle) laser annealing at 0.43 $J/cm^2$/pulse. The $I_{ds}$–$V_{ds}$ curves of as-fabricated devices deviated significantly from the expected response of a long-channel transistor even when $V_{ds}$ values were in the saturation region ($V_{ds} > V_{gs} - V_{th}$), and exhibited significant drain conductance or low output resistance ($r_o$). The annealed devices, on the other hand, appeared to have induced $V_{th}$ shifts to the positive direction, which resulted in smaller saturation current at the same gate voltage. However, the drain currents showed significantly higher output resistance.

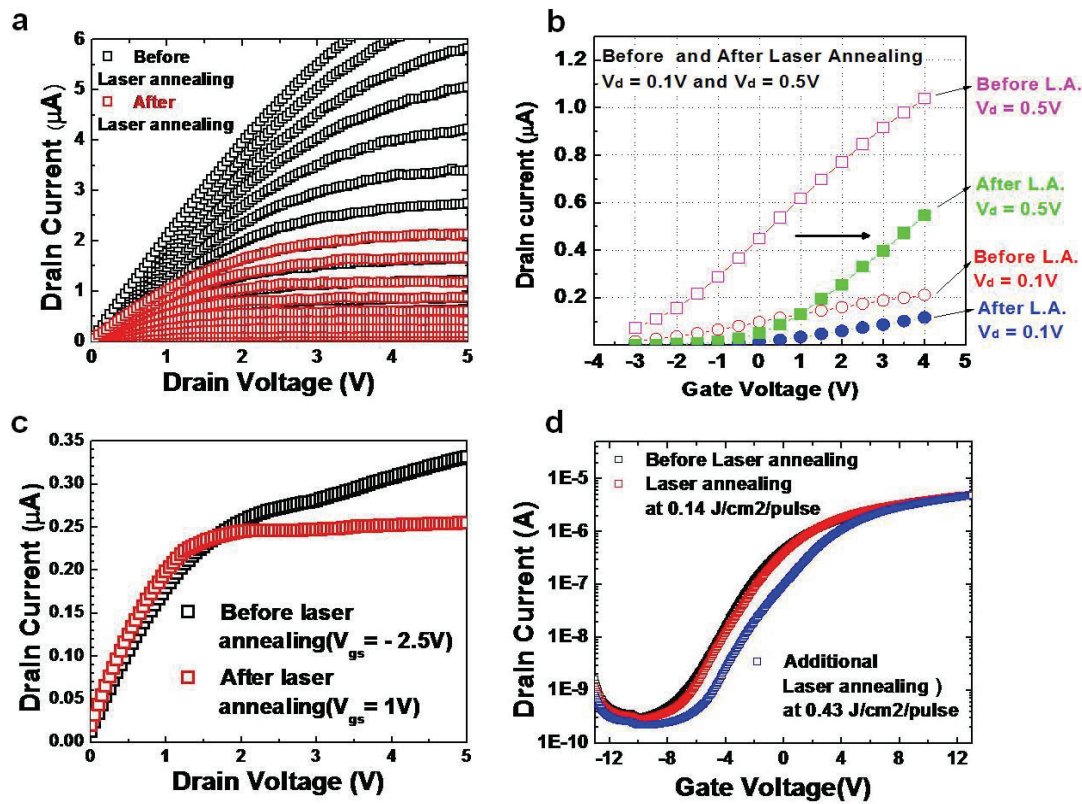

Figure 2 | **Effect of femtosecond laser annealing on the output resistance and threshold voltage of a NWT. a,** The $I_{ds}$–$V_{ds}$ characteristic of a fully transparent $In_2O_3$ NWT. $V_{gs}$ ranges from -1.5 V to 4 V in 0.5 V steps before (black open square) and after (red open circle) laser annealing. **b**, $V_{th}$ shift of the NWT before and after laser annealing at drain-to-source voltages of $V_{ds}$ = 0.1 V and $V_{ds}$ = 0.5 V. **c,** The $I_{ds}$–$V_{ds}$ characteristic for $V_{gs}$ = – 2.5V before the laser annealing (blue curve), and for $V_{gs}$ = 1V after the laser annealing (red curve). The saturation currents are similar, yet the output resistance significantly increased after laser annealing. **d,** the log-scale $I_{ds}$–$V_{ds}$ characteristic of an $In_2O_3$ NWT at $V_{ds}$ = 0.5 V with different power conditions: before applying femtosecond laser annealing (black open square), after 0.14 J/cm$^2$/pulse

femtosecond laser annealing (red open circle) and after an additional 0.43 J/cm$^2$/pulse femtosecond laser annealing (blue open diamond), respectively.

We first identify the threshold voltages before and after the femtosecond laser annealing. The linear-scale drain current versus gate-source voltage ($I_{ds}$–$V_{gs}$) of the fully transparent single $In_2O_3$ NWT at $V_{ds}$ = 0.5 V and $V_{ds}$ = 0.1 V before (square) and after (circle) laser annealing are shown in Fig. 2b. The $V_{th}$ can be extrapolated from the slop of the drain current increase and the values were ~ –2.9 V at $V_{ds}$ = 0.1 V and ~ –2.7 V at $V_{ds}$ = 0.5 V for as-fabricated devices. However, the $V_{th}$ values shifted along positive direction to $V_{th}$ ~ 0.2 V and $V_{th}$ ~ 0.5 V, respectively, after the laser annealing. Data from other $V_{ds}$ values showed similar results and we estimate the threshold voltage to be around –2.8V for as-fabricated NWT and around 0.4 V for annealed NWT. The apparent reduction in source-drain current after the laser annealing can thus be explained by the positive shift of the threshold voltage.

To compare the output resistance, we plotted the $I_{ds}$–$V_{ds}$ characteristics at $V_{gs}$ = –2.5V for the as-fabricated device, and at $V_{gs}$ = 1V for the annealed device (Fig. 2c). The saturation currents were similar, as the $V_{gs}$ – $V_{th}$ were similar, (0.3V for as-fabricated and 0.6V for annealed NWT). For $V_{ds}$ > 1.5V, which is appreciably higher than $V_{gs}$ – $V_{th}$, the device should be in saturation state. However, the as-fabricated device clearly showed a weak saturation, or small output resistance, while the annealed device showed strong saturation. We applied linear regression to calculate the output resistance of the transistor using $I_{ds}$–$V_{ds}$ data in the range of 1.5V < $V_{ds}$ < 5V. The output resistance for the as-fabricated transistor was 37 MΩ, while for the annealed sample it was 200 MΩ, showing a 5.4 fold increase. Similar increase of output resistance (3-7 folds) was observed at other saturation current values. Strong saturation is very important for almost all circuit applications requiring transistors and we believe our method is the first to achieve such a goal with extremely low thermal budget, and without surface modification.

Temporary $V_{th}$ shifts have been reported for $In_2O_3$ NWTs after UV light exposure[20]. However, such exposure shifts the threshold to the negative direction and the device returns to its previous operation state shortly. The effect of femtosecond laser annealing appears to be permanent, and is stable in air. When we re-measured nanowire transistors after a few days and after several weeks, we observed negligible variations.

This permanent change of $V_{th}$ suggests that the post-metallization S/D annealing with a femtosecond laser could also be a tuning method to adjust the $V_{th}$ values of individual nanowires. To illustrate this potential, two different values of annealing power were sequentially applied to the same nanowire transistor and we observed a positive $V_{th}$ shift after each annealing. We first measured the $I_{ds}$–$V_{gs}$ ($V_{ds}$ = 0.5 V) of another representative NWT before laser annealing, and found the $V_{th}$ to be $-1$ V, and then applied femtosecond laser annealing at 0.14 J/cm$^2$/pulse. A $V_{th}$ shift to the positive direction by 0.5 V was observed. We then performed a second annealing on the same device, with the energy of 0.43 J/cm$^2$/pulse. A further shift towards the positive direction by 2.25 V was shown in Fig. 2d. The additional power (in our case 0.43 J/cm$^2$/pulse) was essential because when we tried to apply the same annealing power, a negligible $V_{th}$ shift was observed. Figure 2d shows the log-scale $I_{ds}$~$V_{gs}$ characteristics of an $In_2O_3$ NWT at $V_{ds}$ = 0.5 V for different annealing conditions: before applying femtosecond laser (black open square, $V_{th}$ = $-1$ V, $I_{on}/I_{off}$ ≈ $1.19\times10^4$, SS = 2.2 V/dec, and $\mu_{eff}$ = $1.12\times10^2$ cm$^2$/V·s); after femtosecond laser annealing at pulse energy of 0.14 J/cm$^2$/pulse (red open circle, $V_{th}$ = $-0.5$ V, $I_{on}/I_{off}$ ≈ $1.76\times10^4$, SS = 2.2 V/dec, $\mu_{eff}$ = $1.47\times10^2$ cm$^2$/V·s); and after an additional femtosecond laser annealing at 0.43 J/cm$^2$/pulse (blue open diamond, $V_{th}$ = 1.75 V, $I_{on}/I_{off}$ ≈ $2.23\times10^4$, SS = 2.2 V/dec, $\mu_{eff}$ =

$1.77\times10^2$ cm$^2$/V·s), respectively. After each femtosecond laser annealing, the $I_{on}/I_{off}$ and $\mu_{eff}$ both improved slightly.

Therefore, femtosecond laser annealing apparently have not only improved current saturation (by increasing output resistance by 3-7 fold) but also adjusted threshold voltages of individual In$_2$O$_3$ nanowire transistors. Such effects might provide a solution to one of the long lasting problems in large scale integration of devices made from NWTs: individual trimming of NWT characteristics to match the requirements of functional devices, such as inverters, current mirrors and amplifiers.

**Shifting the switching voltage of a fully transparent inverter**

As an application for our capability of adjusting the $V_{th}$ values of individual NWTs, we fabricated a fully transparent inverter with both transistors made from In$_2$O$_3$ nanowires. An inverter is one of the fundamental building blocks of logic circuits, and its switching threshold (or trip) voltage is preferred to be located at the middle of the supply voltage, which requires the proper positioning of the $V_{th}$ values of both transistors. Moreover, high $r_o$ and early saturation of the transistors are also desirable to improve the noise margin by maintaining the gain in the transition region. Femtosecond laser annealing introduced here appears to be an ideal method to improve the inverter characteristics. Figure 3a shows the two types of inverters we have fabricated, one with depletion mode load (left) and the other with enhanced mode load (right). The two types of inverters are the possible candidates when there is no complementary component such as *p*-type nanowire MOS in the pull-up path. SEM images of depletion mode inverter with the pull up and pull down paths are shown in Fig. 3b. Both topologies worked successfully with a supply voltage of 4 V throughout the experiments. Femtosecond laser annealing was selectively applied to individual transistors to improve the voltage transfer characteristic (VTC) of inverters, specifically the noise margins, which are defined as follows:

$$NM_H = V_{DD} - V_{IH}$$
$$NM_L = V_{IL},$$
[1]

where $V_{IL}$ and $V_{IH}$ are input voltages at the operational points where $\frac{dV_{OUT}}{dV_{IN}} = -1$

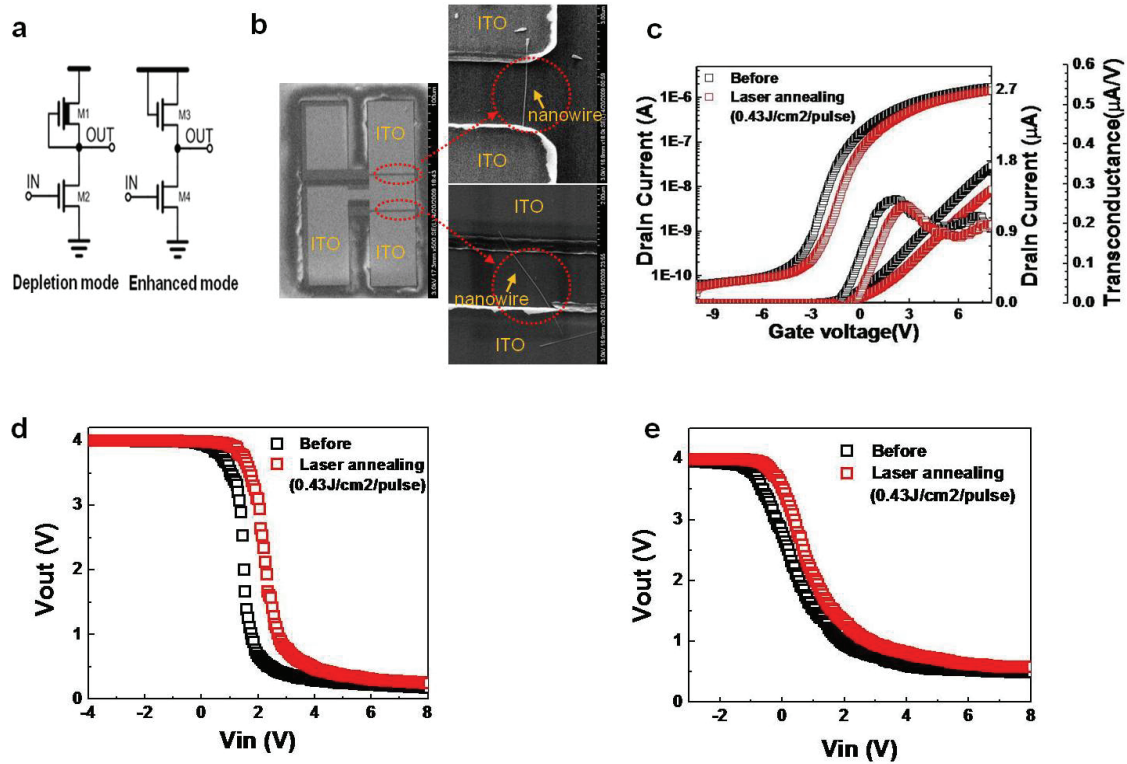

Figure 3 | **Shifting the switching threshold voltage of an inverter consisting of two NMOS NWTs. a,** Schematic for the circuit of depletion (left) and enhancement (right) mode inverters. **b,** SEM images of depletion mode inverters with pull up and pull down path. (c) The drain current versus gate-source voltage ($I_{ds}$-$V_{gs}$) of the fully transparent single $In_2O_3$ NWT at $V_d$ = 0.5 V. The threshold voltage ($V_{th}$), on-off current ratio ($I_{on}$/$I_{off}$), field effect mobilities ($\mu_{eff}$), and subthreshold slope (SS) of NWTs before laser annealing were -0.25 V, ~3 x $10^4$, 83.6 $cm^2V^{-1}S^{-1}$, and ~0.9 V/dec, respectively. After laser annealing with a fluence of 0.43 $J/cm^2$/pulse, those values were changed to 0.6 V, ~3.2 x $10^4$, 78.6 $cm^2V^{-1}S^{-1}$ and ~0.9 V/dec, respectively. **d,** Voltage transfer curves of

the inverter before (black squares) and after (red squares) the laser annealing for the depletion mode load. **e,** Voltage transfer curves before (black squares) and after (red squares) the laser annealing for the enhanced mode load.

$NM_L$ and $NM_H$ represent noise immunity on input logic values: '0' and '1', respectively. Thus, a balance between $NM_L$ and $NM_H$ is required to maximize noise immunity on both logic inputs, and the gain by the inverter in the transition region has to be maintained high to preserve the total noise margin ($NM_L + NM_H$). As shown in Fig. 3c, the laser annealing maintained transconductance (changes were insignificant) of NWT while it shifted $V_{th}$. This allowed us to control switching threshold voltage of an inverter with the same gain at the switching threshold voltage ($V_M$), or trip voltage, which will maximize noise margin of the inverter. In the case of depletion mode inverter, the diode connected NMOS (M1) is always ON as M1 has a negative $V_{th1}$ and its $V_{gs1}$ is fixed at 0, see Fig. 3a. When the input is low ('0') and transistor M2 is off, M1 keeps driving the output high until $V_{sd}$ of M1 drops to zero, which means that $V_{OUT}$ is the same as supply voltage. When the input state changes to high ('1'), M2 starts to discharge output quickly. This can be explained by the relative magnitude of $V_{gs} - V_{th}$ for M2 and for M1, which is fixed at $-V_{th1}$ since $V_{gs1}$ for M1 is always 0. When $V_{gs2} - V_{th2} = V_{IN} - V_{th2}$ for M2 is larger than $-V_{th1}$ of M1, the current is limited by M1; and $V_{ds2}$ of M2 quickly reduces to near zero to match the small current set by M1. This ensures a fast switching from high to low. Therefore the trip voltage is mostly determined by the $V_{th}$ of M2 and $r_o$ of M1 and M2, and could be smaller (1.5 V) than half of the supply voltage, 2V, as shown in Fig. 3d. To achieve enhanced noise margin, the trip voltage is preferred to be shifted to close to 2 V. Moreover, the function of M1 should remain complementary to that of M2, so the threshold voltage of M1 had to be maintained negative while that of M2 is shifted along positive direction. This requires local tuning of the pull down transistor (M2) without significantly affecting the pull-up transistor (M1). Our femtosecond laser annealing meets those requirements and can be applied selectively to the pull down transistor to shift the switching voltage of inverter to be in the middle of the supply rail. The voltage transfer characteristics in Fig. 3d show that enhanced noise margin was achieved by shifting the trip point of inverter from 1.5 V to 2.2 V. Thus, it is possible to use this technique to control switching threshold voltage of an inverter, which is important to achieve a high noise margin for many circuit applications.

The operating principle of enhancement mode load transistor is different compared to depletion mode load inverter. Figure 3(e) shows that output voltage was not completely zero even when the input was driven high. Also the transition from high to low was not as sharp as that of the depletion mode. These were primarily due to the static current through M3 and M4 when M4 was turned on. Unlike the depletion mode, the $V_{gs3} - V_{th3}$ increases when $V_{OUT}$ drops, which increases the static current. At this time, the output voltage was determined by the on resistance ($R_{ON}$) values of M3 and M4 as Ohm's law is applicable. Thus, the ratio of pull up and pull down transistor was important in this case. In general, this ratio can be achieved by adjusting the channel length. In addition, high $R_{ON}$ of M3 was required to obtain a sharper transfer from high to low state. Note that both transistors were in the saturation mode in the middle of transition. Figure 3e shows that laser annealing can also improve the noise margin by shifting the trip voltage to the positive direction toward half of the supply voltage.

Finally, our inverter is highly transparent. Figure 4 shows the optical transmission spectra through the fully transparent NMOS inverters using $In_2O_3$ nanowires on a glass substrate in the 350-1250 nm wavelength range. The optical transmission value was ~ 82%. Note that the optical transmission value of corning glass substrate is ~ 92%. The NWT array regions were 0.5 x 0.1 inch (the glass substrate was 1.5 x 1.0 inch) and contained ~1500 NWT device patterns; and the entire substrate was coated with the $Al_2O_3$ gate insulator. The source/drain regions and the gate regions covered ~40% and

~60% of the total NWT array region, respectively. Since $In_2O_3$ nanowires do not cover much of entire NWT array and the diameter of the NWs was only 20nm, their optical absorption was negligible. The inset in Figure 4 shows the substrate with fully transparent NMOS inverters over an opaque layer. The texture on the paper is clearly seen through the device substrate.

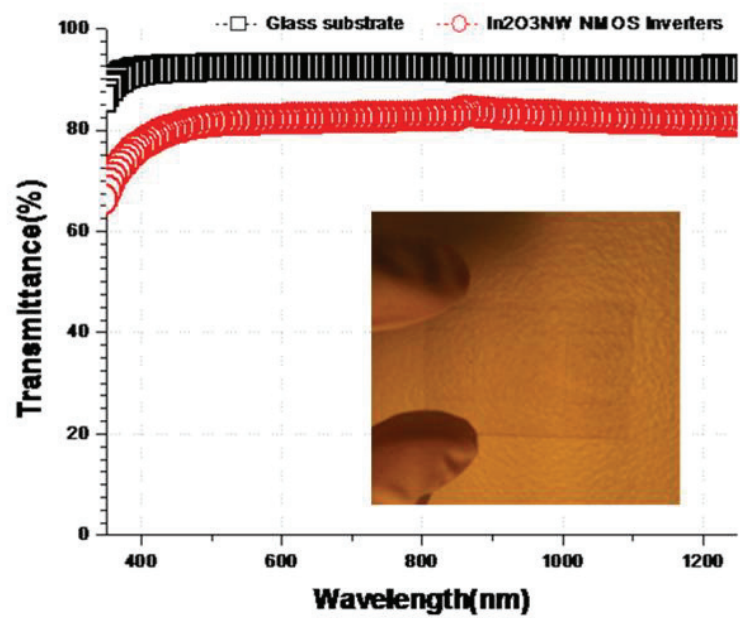

Figure 4 | **Optical transmission spectrum through the entire NWT inverter structures.** The inset shows the high transparency of the substrate with 1,500 NWT inverter devices; with the texture on the layer below the substrate clearly visible.

**Discussion**

It is important to improve the performance of as-fabricated nanowire devices as they typically suffer from weak saturation and unpredictable threshold voltages. The thermal budget of annealing is typically limited by the low-temperature requirements of transparent and flexible substrates. Femtosecond lasers could be focused and tune individual NWTs. However, they can also damage the NWTs easily. The direct illumination of nanowires was avoided in our annealing process so that damaging of NWTs did not occur. This was evidenced by the preservation and slight improvement of other major performance parameters, such as mobility, on-off current ratio, and subthreshold slope. The improvement of current saturation, on the other hand, is desirable in most applications.

Since our femtosecond laser photons have energy below the band gap of $In_2O_3$ nanowires, femtosecond laser annealing is expected to be mainly thermal, possibly forming an improved single-crystalline $In_2O_3$ nanowire structure. The short pulse duration may result in ITO photo-physical bond breaking instead of classical melting[21], consequently forming ITO spikes into the nanowire channel to improve the contact-channel interface, modifying the Schottky barrier height and the effective doping in the nearby semiconductor region. Further investigation of the mechanism behind such annealing effects is interesting and ongoing. This study provides insights into the contact-dominated transistor properties, in terms of the effects on output resistance and $V_{th}$.

Combined with the excimer laser annealing[13], which shifts the threshold voltage to the negative direction by increasing the number of oxygen vacancies, one could envision full trimming capability of the threshold voltages of NWTs and maintaining high current saturation, thus opening the possibility of constructing sophisticated circuit blocks or other functional devices made from NWTs, and significantly advance our knowledge on flexible, and transparent electronics on low-temperature substrates. Controlling the threshold voltages of nanowires is of central importance to any practical integrated circuits. The semiconductor industry enjoys highly uniform doping and high-precision manufacturing (i.e. critical dimension control) to achieve uniform threshold voltages. While manufacturing of non-Si nanowire based transistors will certainly improve with novel techniques, it is unlikely that they will match the level of control in CMOS technologies, therefore the femtosecond laser tuning of individual NWT presented here would be very important in manufacturing NWTs if large circuit blocks are to function as designed.

We note that there could be other ways to alter the transistor characteristics, such as surface passivation and chemical modifications. Femtosecond laser annealing appears to be non-invasive, and still preserve the flexibility of applying the abovementioned tuning process. Thus it would be a useful trimming method for future NWT based integrated circuit manufacturing.

**Acknowledgements** We thank Dr. Jun Liu and Prof. T. J. Marks for help on depositing ITO films. This research was supported by the Defense Advanced Research Projects Agency under contract N66001-08-1-2037, by the National Science Foundation (NSF) under contract NIRT-0707817, and by the by the Basic Science Research Program and the Converging Research Centre Program through the National Research Foundation of Korea (NRF) funded by the Ministry of Education, Science and Technology (2009-0057214 & 2009-0082818).

**Author Contributions** C. L. fabricated the devices, performed electrical characterization, and processed the data; P. S. performed femtosecond laser annealing; C.Z provided nanowires; S. P., S. K, D. B. J., X. X., K. R., S. J. and M. Q. discussed experimental procedure and results. S. J. designed the layout, and performed early annealing experiments. M. Q. proposed the idea. C. L., S. J. and M. Q. wrote the manuscript.

**Competing interests statement** The authors declare that they have no competing financial interests.

**Correspondence** and requests for materials should be addressed to M. Q. (mqi@purdue.edu) and S. J. (shju@kgu.ac.kr).

## Methods

### Fabrication

30nm of high-k $Al_2O_3$ was deposited by a custom ALD system on corning 1737 glass substrates. Individual gate electrodes were formed by sputter deposition of ITO ($R_{sheet}$ = 60 ohms per square) followed by photolithography patterning. Single-crystalline semiconducting $In_2O_3$ nanowires were synthesized by a pulsed laser ablation process, and the average diameter and length of the nanowires are ~20 nm and ~5 μm, respectively. $In_2O_3$ nanowires were then dispersed in a solution which forms a puddle on top of the substrate with patterned gates. After the solution evaporated, some of the nanowires were deposited on gate regions, and ITO source and drain electrodes were selectively deposited by sputtering method. Following the source-drain patterning and electrical characterization, femtosecond laser anneals were performed at the source/drain regions.

**Laser annealing procedure**

Laser annealing source was a Ti:Sapphire laser operating at 800 nm. The pulse duration was 50 fs, and the pulse repetition rate was 1 kHz. A three-axis computer controlled positioning stage was used to move the sample with respect to the laser beam. The laser machining system was equipped with an in-line vision system which allowed laser annealing at the desired locations on the devices. The laser pulses were focused through a 100× microscope objective lens with a numerical aperture of 0.8, which produced a spot of 1.22 μm in diameter on the target surface. The scanning speed was 1 μm/s. Two different energy fluences, 0.14 J/cm²/pulse, and 0.43 J/cm²/pulse (corresponding to average pulse energy of 1.67μW and 5μW, respectively), were used in sequence on a selected nanowire transistor.

**Optical transmission measurement**

The transmission spectra of normal incident linearly polarized light were collected using a Lambda 950 spectrophotometer (Perkin-Elmer). Electrical characterizations was performed using a semiconductor parameter analyzer (HP 4156A).

**Calculation of the field-effect mobility**

The field-effect mobility $\mu_{eff} = \dfrac{\Delta I_{ds} L^2}{\Delta V_{gs} C_i V_{ds}}$ was calculated using the cylinder-on-plate (COP) capacitance model $C_i = \dfrac{2\pi\varepsilon k_{eff} L}{\cosh^{-1}\left(1+\dfrac{t_{ox}}{r}\right)}$.